\documentclass[aps,prb,twocolumn, superscriptaddress, floatfix, floats, amssymb, amsmath, showpacs]{revtex4}
\usepackage{calc}
\usepackage{psfrag}
\usepackage{graphicx}
\usepackage{color}
\begin{document}
\def \beq{\begin{equation}}
\def \eeq{\end{equation}}
 \newcommand{\sgn}{\mathop{\mathrm{sgn}}}

\newcommand{\beqn}{\begin{eqnarray}}
\newcommand{\eeqn}{\end{eqnarray}}
\newcommand{\nn}{\nonumber}

\newcommand{\ua}{\uparrow}
\newcommand{\da}{\downarrow}

\newcommand{\rhobar}{\bar{\rho}}
\newcommand{\Sbar}{\bar{S}}

\newcommand{\bq}{{\bf q}}
\newcommand{\bR}{{\bf R}}

\newcommand{\red}{\color{red}}

\title{Fractional quantum Hall state at $\nu=1/4$ in a wide quantum well}

\author{Z. Papi\'{c}}
\affiliation{Laboratoire de Physique des Solides, Univ.~Paris-Sud, CNRS, UMR 8502, F-91405 Orsay Cedex, France}
\affiliation{Institute of Physics, P.~O.~Box 68, 11 000 Belgrade, Serbia}

\author{G. M\"{o}ller}
\affiliation{Theory of Condensed Matter Group, Cavendish Laboratory, J.~J.~Thomson Ave., Cambridge CB3 0HE, UK}

\author{M. V. Milovanovi\'c}
\affiliation{Institute of Physics, P.~O.~Box 68, 11 000 Belgrade, Serbia}

\author{N. Regnault}
\affiliation{Laboratoire Pierre Aigrain, Ecole Normale Sup\'erieure, CNRS, 24 rue Lhomond, F-75005 Paris, France}

\author{M. O. Goerbig}
\affiliation{Laboratoire de Physique des Solides, Univ.~Paris-Sud, CNRS, UMR 8502, F-91405 Orsay Cedex, France}

\date{\today}

\begin{abstract}
We investigate, with the help of Monte-Carlo and exact-diagonalization calculations in the spherical geometry, several compressible and incompressible candidate wave functions for the recently observed quantum Hall state at the filling factor $\nu=1/4$ in a wide quantum well. The quantum well is modeled as a two-component system by retaining its two lowest subbands. We make a direct connection with the phenomenological effective-bilayer model, which is commonly used in the description of a wide quantum well, and we compare our findings with the established results at $\nu=1/2$ in the lowest Landau level. At $\nu=1/4$, the overlap calculations for the Halperin (5,5,3) and (7,7,1) states, the generalized Haldane-Rezayi state and the Moore-Read Pfaffian, suggest that the incompressible state is likely to be realized in the interplay between the Halperin (5,5,3) state and the Moore-Read Pfaffian. Our numerics shows the latter to be very susceptible to changes in the interaction coefficients, thus indicating that the observed state is of multicomponent nature.
\end{abstract}


\maketitle \vskip2pc

\section{Introduction}

Advances in fabrication of high-quality GaAs semiconductor systems have led to an ever growing collection of the observed incompressible fractional quantum Hall states
in a variety of settings.\cite{pan} These states occur at particular ratios between the number of electrons $N$
and the number of magnetic flux quanta $N_\phi$ that pierce the system in the direction perpendicular to the sample. This commensurability can be expressed as 
the filling factor $\nu=N/N_\phi=p/q$ in terms of integers $p, q$, which is the single most important quantity that characterizes the quantum Hall state. 

In a thin layer, $q$ usually turns out to be an odd integer, the fact which
had its pioneering explanation in terms of the Laughlin wave function\cite{laughlin} for the case of $p=1, q=3, 5, 7, ... $ and its subsequent generalizations in terms of composite
fermions\cite{jain}  (CF), applicable to general integers $p, q$ as long as $q$ is odd, and hierarchy theory.\cite{hierarchy}
However, a state with an even denominator has also been observed\cite{willet} but in the first excited Landau level (LL).  One cannot account for it in the usual Laughlin/composite fermion approach and the idea of pairing has commonly been invoked to explain the origin of this fraction.\cite{mr, hr} The simplest realization of pairing between spin polarized electrons is the so-called Pfaffian defined by the Moore-Read wave function\cite{mr} and supporting excitations with non-Abelian statistics.\cite{readrezayi}

The possibility of an extra degree of freedom lifts the requirement of Fermi antisymmetry and hence gives another route towards realizing even denominator fractions. The additional degree of freedom can be the ordinary spin or else a ``pseudospin" in case of a wide quantum well, where the two lowest electronic subbands correspond to $\uparrow, \downarrow$. If the sample is etched in such a way to create a barrier in the middle, thus supressing tunneling between the two ``sides", one can think of it as a bilayer with $\uparrow, \downarrow$ denoting the left and right layer where electrons can be localized.  Incompressible quantum Hall states for such systems have been theoretically predicted in Ref.~\onlinecite{hraps}  and experimentally confirmed for cases of bilayer at filling factor $\nu=1$ and $\nu=1/2$.\cite{eisen1/2, suen_bilayer}  Later on, essentially the same quantum Hall state at $\nu=1/2$ was observed in a sample which had the geometry of a single wide well.\cite{suen} It was argued, on the basis of a self-consistent Hartree-Fock approximation, that in a wide well the electrons (due to their mutual repulsion) reorganize themselves  so as to form an effective bilayer distribution of charge. Hence, an equivalence between the two very different samples was claimed and theoretical works set out to analyze the problem from this premise.\cite{he, y2} 

On the basis of the quantum mechanical overlap with the ground state obtained in exact diagonalization (ED), including a realistic bilayer confinement potential, Ref.~\onlinecite{he} established that the ground state is well described by the so-called (3,3,1) Halperin wave function.\cite{halperin} This wave function distinguishes between two kinds of electrons and the fact that it describes the system is what we mean by the system being ``multicomponent". Experimental work gave further insight into the nature of the multicomponent state at $\nu=1/2$ and strengthened the belief that the (3,3,1) wave function is a correct physical description.\cite{suen} Namely, the behavior of the excitation gap as a function of tunneling amplitude $\Delta_{SAS}$ (i.e. the splitting between the two lowest subbands) was found to have upward cusp at the intermediate value of $\Delta_{SAS}$ and the state was quickly destroyed by the application of electrostatic bias (charge imbalance).\cite{suen} In Ref.~\onlinecite{y2}, a numerical study was able to reproduce the observed upward cusp in the activation gap by diagonalizing the bilayer Hamiltonian with explicit interlayer tunneling. 

A recent experimental paper\cite{luhman} reports the observation of the $\nu=1/4$ quantum Hall state in a wide quantum well. The state is fragile and almost indiscernible when only a perpendicular magnetic field is applied (although one could expect that with yet higher sample qualities, a small plateau would be developed already at that point).  However, when the magnetic field is tilted, there is a clear dip in the value of longitudinal resistance $R_{xx}$, signifying the presence of an incompressible state. 

In this paper we analyze the complex interplay between the single- and multicomponent nature of the ground state at $\nu=1/4$ in a wide quantum well, in comparison with the ground state at $\nu=1/2$. Contrary to previous studies,\cite{he, y2} we do not make the {\sl ad hoc} assumption that the wide quantum well may be described as an effective bilayer. Instead, we consider the two lowest electronic subbands of the quantum well, which is modeled by the infinite square well for the sake of convenience but cross-checked with other confinement models. The energy splitting between these two subbands, the associated wave functions of which are symmetric and antisymmetric, respectively, in the $z$-direction, is given by $\Delta_{SAS}$ (occasionally referred to as the tunneling amplitude). Due to the low filling factor ($\nu=1/4$), the power of the ED method will be rather limited and other complementary approaches may be needed to fully explain the experimental findings.

The paper is organized as follows. 
Sec.~\ref{pfaffian_section} is devoted to the single-component candidate for $\nu=1/4$, and we study its overlap with the exact Coulomb ground state within various confinement models. In Sec.~\ref{twocomponent}, we define the multicomponent wave functions expected to be relevant at this filling factor. The two likely candidates, the Halperin (5,5,3) and (7,7,1) states, are investigated within a simple bilayer model without tunneling. The two-subband model of the quantum well is introduced and described in Sec.~\ref{quantumwell_section}. Our main results of ED calculations in the spherical geometry are presented in Sec.~\ref{section_competingphases}.  To extend the reach of our numerics, we furthermore deploy Monte-Carlo simulations of the trial wavefunctions identified beforehand to analyze their energetic competition. We summarize with our view on the nature of the state at $\nu=1/4$ in Sec.~\ref{conclusion_section}.

\section{One component state}\label{pfaffian_section}

\subsection{Pfaffian at $\nu=1/4$} \label{pfaffian_definition}

There is a natural candidate for the fully polarized quantum Hall state at $\nu=1/4$ -- it is the generalized Moore-Read Pfaffian:\cite{readrezayi}

\beq \label{pfaffian}
\Psi_{\rm Pf}(z_1, ..., z_N)=\mathrm{Pf}\left(\frac{1}{z_i  - z_j}\right) \prod_{i<j} (z_i - z_j)^4,
\eeq
expressed in terms of the complex coordinate of the electron in the plane where $z_j=x_j+iy_j$. The object $\mathrm{Pf}$ is defined as 
\begin{equation}
\nonumber \mathrm{Pf}M_{ij}=\frac{1}{2^{N/2}(N/2)!}\sum_{\sigma \in S_N} \sgn \sigma \prod_{k=1}^{N/2}M_{\sigma (2k-1) \sigma (2k)},
\end{equation}
acting upon the antisymmetric $N \times N$ matrix $M_{ij}$, and $S_N$ is a group of permutations of $N$ objects. $\mathrm{Pf}$ renders the wave function totally antisymmetric and encodes the same kind of correlations as in the more familiar $\nu=5/2$ case.\cite{mr} In the spherical geometry\cite{hierarchy, haldane_rezayi_ed} many-body states are characterized by the number of electrons $N$, the number of flux quanta $N_{\phi}$ generated by a magnetic monopole placed in the center of the sphere and extending radially through its surface, and an additional topological number which is the shift. For the Pfaffian in Eq.~(\ref{pfaffian}), the three numbers are related by the formula $N_{\phi}=4N-5$. $\Psi_{\rm Pf}$ is a zero-energy eigenstate of a certain 3-body Hamiltonian,\cite{readrezayi} but in our calculations it was generated from its root configuration via the squeezing technique.\cite{jack} On the other hand, the Coulomb (two-body) Hamiltonian commutes with the angular momentum operator $\mathbf{L}$ because of rotational invariance and, by Wigner-Eckart theorem, the interaction is parametrized by discrete set of numbers $V_L$ known as the Haldane pseudopotentials.\cite{hierarchy} The motion of electrons is therefore fully described in terms of the in-plane (spherical) coordinates $\theta, \phi$ and the use of different confinement models in the (perpendicular) $z$-direction (neglecting the in-plane magnetic field) will only modify the values of pseudopotentials. 

\subsection{Finite thickness models} \label{finite_thickness}

Most of the candidate wave functions for quantum Hall fractions have been extensively studied via numerical techniques such as ED or Monte Carlo. For the sake of convenience, but also due to the intrinsic ambiguity which stems from the fact that in a strongly correlated system many input parameters (e.g. the precise form of the interaction) are unknown, it is natural to start off from the limit of infinitely thin layer of electrons interacting via Coulomb force and hope that the inclusion of e.g. realistic confinement and sample thickness will have small, perturbative corrections. There have been different proposals to account for the finite thickness of the sample in the perpendicular direction, but the one that is straightforward and most natural from the point of view of ED is the Zhang-Das Sarma (ZDS) model\cite{zds} which is simply given by substituting the interaction
\begin{eqnarray}\label{zds_model}
\frac{1}{r} \rightarrow \frac{1}{\sqrt{r^2+(w/2)^2}}
\end{eqnarray} 
(we will always denote by $w$ the width of the sample and the energy is always expressed in units of $e^2/\epsilon l_B$, where the magnetic length is $l_B=\sqrt{\hbar c/eB}$ is given in terms of the perpendicular magnetic field $B$).  Qualitatively, this substitution softens the interaction\cite{zds} and was studied  extensively (together with other confinement models, some of which we will introduce below) in Ref.~\onlinecite{pjds}, where it was advertised to significantly stabilize the Moore-Read Pfaffian at $\nu=1/2$ (the effect being most pronounced in the second LL), but (in most cases) decrease the overlap somewhat for the Laughlin states at $\nu=1/3$ and $1/5$. In Ref.~\onlinecite{park} it was noticed that this kind of interaction can lead to an instability of the composite fermion sea, which is believed to describe the compressible state at $\nu=1/2$ in the lowest LL, towards the paired state described by the Pfaffian. Indeed, the CF Fermi liquid can be regarded as a special member of the general class of paired CF wavefunctions,\cite{ms08} of which it represents the limit of vanishing gap.

Although the ZDS model (\ref{zds_model}) has a very simple form, there is no physical wave function that corresponds to this confinement potential in the $z$-direction. Other popular choices for the confinement in the $z$-direction include the infinite square well (ISQW) and Fang-Howard (FH), which are presumably more realistic than ZDS because they are defined by the actual wave functions of simple model potentials for the quantum well, given by

\begin{eqnarray}
\phi_{\rm ISQW}(z) &=& \sqrt{\frac{2}{w}}\sin\left(\frac{\pi z}{w}\right), \label{isqw}\\
\phi_{\rm FH}(z) &=& \sqrt{\frac{27}{2w^3}}\, z\,  e^{-3 z/2w}, \label{fh}
\end{eqnarray}
respectively. 

\subsection{Overlaps}\label{pfaffian_overlaps}

\begin{figure}[ttt]
\centering
\includegraphics[width=\linewidth, angle=270, scale=0.7]{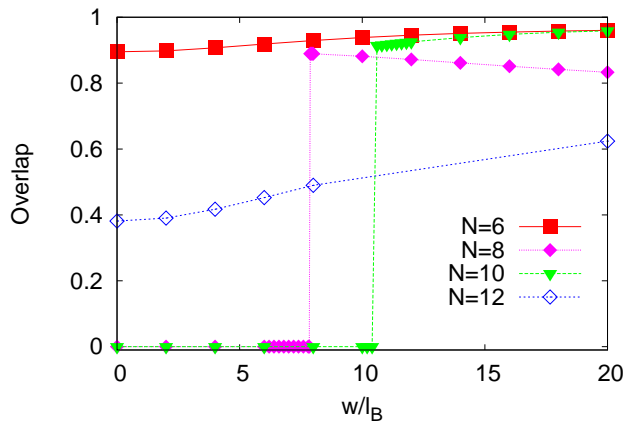}
\caption{(color online) Overlap $|\langle \Psi_{\rm Pf}|\Psi_{\rm exact}\rangle|$ between the exact Coulomb state for finite width (ZDS model) and the Pfaffian at $\nu=1/4$.}\label{pfaffian_width_overlap}
\end{figure}

We have performed ED calculations for various confinement models (\ref{zds_model}-\ref{fh}) and all system sizes $N=6, 8, 10, 12$ accessible at present. In Fig.~\ref{pfaffian_width_overlap} we present the overlap $|\langle \Psi_{\rm Pf}|\Psi_{\rm exact}\rangle|$ between the exact Coulomb ground state at $\nu=1/4$ and $\Psi_{\rm Pf}$, finite width being modeled by the ZDS ansatz (\ref{zds_model}). The size of the Hilbert space at $N=12$ is noteworthy: the dimension of the $L_z =0$ sector is $218\,635\,791$.

It appears that the overlap of the Pfaffian state is rather high for large values of the width (even if it is negligible for small $w$'s). These values could likely be increased further by considering general pairing wave functions.\cite{ms08} However, these overlaps alone cannot be taken as solid evidence for a pairing nature of the $\nu=1/4$ for two reasons. First, for $N=6$ and $N=12$ there is the aliasing problem with composite fermion states: Jain states with different physical properties (e.g. Abelian instead of non-Abelian statistics) occur at the same values of $N$ and  $N_{\phi}$ on the sphere (because of finite system size). 
High overlap for the aliased states may therefore come from other incompressible states different from the Pfaffian. Secondly, for the non-aliased states at $N=8$ and $N=10$, there appears to be a critical value of the width at which the overlap as a function of $w$ suffers a sharp jump. By analyzing the entire low energy spectrum on the sphere as a function of width, we have established that the (neutral) gap collapses at the critical point of $w/l_B$. Therefore, in order to get to the Pfaffian phase, one must go through a (first order) phase transition. Before the transition, the ground state is obtained in the $L>0$ sector of the Hilbert space and the overlap with the Pfaffian (which resides in $L=0$ sector) remains zero due to the difference in symmetry.

The lack of adiabatic continuity and the aliasing problem cast some doubt on the Pfaffian state as a good candidate for $\nu=1/4$ in the lowest LL. We have also checked using other confinement models (\ref{isqw}, \ref{fh}), but in these cases for $N=8$ and $N=10$ the overlap remains zero for any value of $w/l_B$.  Thus our ED results do not yield a definite answer with respect to the relevance of $\Psi_{\rm Pf}$ in the single layer at $\nu=1/4$. 

We would like to stress the qualitative difference in our results obtained by using ZDS versus other confinement models which appears, to the best of our knowledge, to be the first such case in the literature. The smaller overall energy scale (and the smaller gap as well) is very likely to be at the origin of this discrepancy. We note in passing that, contrary to the finite-width models which change \emph{all} pseudopotentials at once, one may start from the pure Coulomb interaction and vary just a few strongest pseudopotentials.\cite{rh} We have tried varying both $V_1$ and $V_3$ , but this procedure does not stabilize the Pfaffian phase in any finite region of the parameter space for $N=8$.

\section{Two-component states}\label{twocomponent}

Soon after Laughlin's wave function describing the incompressible state at $\nu=1/3$ when the electron spins are fully polarized, Halperin\cite{halperin} proposed a class of generalized wave functions defined as

\begin{eqnarray}\label{halperinwf}
\nonumber \Psi_{mm'n} (z_{1}^{\uparrow}, ..., z_{N_{\uparrow}}^{\uparrow}, z_{1}^{\downarrow}, ..., z_{N_{\downarrow}}^{\downarrow})  \\
= \prod_{i<j}^{N_{\uparrow}} (z_i^{\uparrow}-z_{j}^{\uparrow})^m \prod_{k<l}^{N_{\downarrow}} (z_{k}^{\downarrow}-z_{l}^{\downarrow})^{m'} \prod_{s}^{N_{\uparrow}} \prod_{t}^{N_\downarrow}(z_s^{\uparrow}-z_{t}^{\downarrow})^n ,
\end{eqnarray} 
where the electrons are distributed over two components (labeled by $\uparrow, \downarrow$). The exponents $m, m'$ denote the ``intra"-component correlations originating from the basic Laughlin-Jastrow building blocks within each component, whereas $n$ describes ``inter"-component correlations (we have omitted the ubiquitous Gaussian factors and implicitly assume that there is a spinor part to this wave function as well as an overall antisymmetrization between $\uparrow$ and $\downarrow$). In order for these wave functions to be eligible candidates for the ground state of the system, one must enforce an additional requirement that they be eigenstates of the Casimir operator of the SU(2) group, i.e. the total spin $\mathbf{S}^2$, as long as the interaction is symmetric with respect to ``intra" and ``inter" components (e.g. the usual case of electrons with spin). 
However, apart from electrons with spin, the wave functions (\ref{halperinwf}) have also been used in bilayer systems where this symmetry is broken as soon as the layer separation is non-zero. In this case, the wave functions (\ref{halperinwf}) need not be eigenstates of the total spin. There have been generalizations of these wave functions in the physics of bilayer systems at total filling factor\cite{srm,msr08,msr09} $\nu=1$ and to more than two components,\cite{ggr1} where further constraints on the possible values of $m,m',n$ were derived within the plasma analogy.\cite{ggr} In a two-component case, these turn out to be the intuitive requirement that ``intra"-component interactions are stronger than ``inter"-component interactions: $m,m' \geq n$. 
For the particular case of two components and $m=m'=n+2$ (which includes $\Psi_{331}$ and $\Psi_{553}$), the  Halperin wave function (\ref{halperinwf}) can be analytically cast into a paired form\cite{rgj,msr09} via Cauchy determinant identity (up to the unimportant phase factor),
\begin{eqnarray}
\nonumber \frac{\prod_{i<j}^{N_\uparrow} (z_i^\uparrow - z_j^\uparrow)\prod_{k<l}^{N_\downarrow} (z_k^\downarrow - z_l^\downarrow)}{\prod_{s}^{N_\uparrow}\prod_{t}^{N_\downarrow} (z_s^\uparrow - z_t^\downarrow)} = \det \left[ \frac{1}{z_i^\uparrow - z_j^\downarrow} \right],
\end{eqnarray}
where the pairing function is given by $\det\left[\frac{1}{z_{i}^\uparrow - z_{j}^\downarrow}\right]$.  In the case of the 111-state, this pairing nature was recently exploited to make a connection to paired composite fermion states, and to construct wave functions interpolating between these two regimes.\cite{msr09} Halperin wave functions are the exact zero-energy eigenstates of the two-body Hamiltonian
\begin{eqnarray}\label{halperinham}
\nonumber H &=& \sum_{i<j} \left[ \sum_{L=0}^{m-1} V_L^{\uparrow \uparrow} P_{ij}^{\uparrow \uparrow}(N_{\phi}-L) 
+ \sum_{L=0}^{m'-1} V_L ^{\downarrow  \downarrow}P_{ij}^{\downarrow \downarrow}(N_{\phi}-L) \right]\\
&& + \sum_{i,j} \sum_{L=0}^{n-1} V_L^{\uparrow \downarrow} P_{ij}^{\uparrow \downarrow} (N_{\phi}-L),
\end{eqnarray} 
where $P_{ij}^{\sigma \sigma'}(L)$ projects onto the state with angular momentum $L$ of particles $i$ and $j$ with respective (pseudo)spins $\sigma$ and $\sigma'$. Besides offering great convenience for handling Halperin wave functions (\ref{halperinwf}) in ED, Eq.~(\ref{halperinham}) enabled counting of the number of excited quasihole states and reaffirming the idea that the states described by Eq.~(\ref{halperinwf}) possess Abelian statistics.\cite{readrezayi}

\begin{figure}[ttt]
\centering
\includegraphics[angle=270,scale=0.7, width=\linewidth]{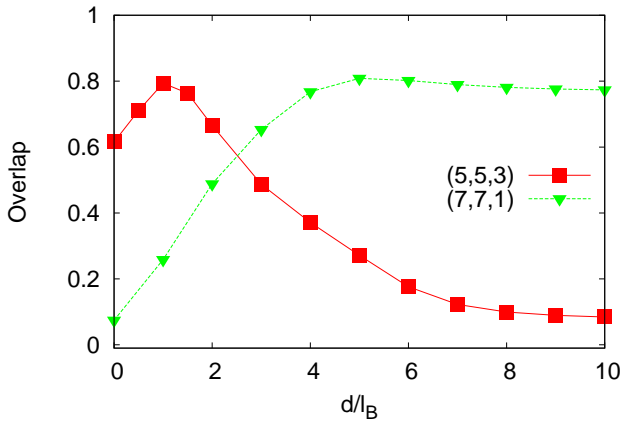}
\caption{(color online) Overlap between the exact bilayer state with (5,5,3) and (7,7,1) states for $N=8$ particles.}\label{553_771}
\end{figure}

At the filling factor $\nu=1/4$, there are three wave functions of the form (\ref{halperinwf}) that meet the necessary physical requirements, $\Psi_{5\,5\,3} \equiv (5,5,3)$, $\Psi_{7\,7\,1} \equiv (7,7,1)$ and $\Psi_{5\,13\,1}\equiv (5,13,1)$. None of them is an eigenstate of $\mathbf{S}^2$, so they are more adapted to the case of a bilayer than that of real spin. In Fig.~\ref{553_771} we present the basic overlap characterization of the first two wave functions in a simple bilayer model defined by the interaction $V^{\uparrow \uparrow}(r)=V^{\downarrow \downarrow}(r)=1/r, V^{\uparrow \downarrow}(r)=1/\sqrt{r^2+d^2}$ ($d$ being the distance between the layers).\cite{ymg} 
$(5,5,3)$ displays a familiar maximum in the overlap for small distance between the layers. $(7,7,1)$ was dismissed in Ref.~\onlinecite{luhman} arguing that it would more likely lead to two coupled Wigner crystals than an incompressible liquid. Our diagonalization scheme is not adapted to address states with broken translation symmetry, so we do not see an \emph{a priori} reason to reject this state. The results in Fig.~\ref{553_771} are for $N=8$ particles, they are fully consistent with those of smaller $N$, but direct comparison between $(5,5,3)$ and $(7,7,1)$ is not possible because they are characterized by different shifts ($-5$ and $-7$ respectively).\cite{ggr} We will address this issue below by extrapolating to the thermodynamic limit the respective trial energies from Monte-Carlo simulations for both of these states.

The last possibility, $(5,13,1)$, is a peculiar one because it can only occur in the case of a strong density imbalance.  Such an imbalance would lead to an increase in the charging energy, but if one of the coupled states is a prominent quantum Hall state, the gain in correlation energy can outweigh the price of charge imbalance, as it has been experimentally verified.\cite{manoh} However, in the present case, our numerical calculations
confirmed that this candidate can be discarded because it takes unrealistically high values of the sample width for this wave function to have any numerical relevance at all.

Given the low filling factor $\nu=1/4$ we are studying, one must also consider the possibility of nearby compressible states that can intervene for some values of the external parameters. Apart from the obvious metallic state similar to the Fermi-liquid-like state proposed by Rezayi and Read,\cite{rr} there is in principle also the Haldane-Rezayi (HR) state,\cite{hr, readrezayi} which is defined by

\beq
\nonumber \Psi_{\rm HR}(\{z_i^\uparrow, z_i^\downarrow\}) 
= \det \left[ \frac{1}{(z_i^\uparrow-z_j^\downarrow)^2} \right] \prod_{i<j}^N (z_i-z_j)^4.
\eeq\label{hr}
The last term is a global Laughlin-Jastrow factor for all particles regardless of their spin. $\Psi_{\rm HR}$ is the zero energy eigenstate of the interaction parametrized by the set of pseudopotentials $V_L=\{ 1, 1, 0, 1, 0,  ...\}$ and occurs at the shift of $-6$.  It is also a spin singlet\cite{hr} and compressible on the basis of its nonunitary parent conformal field theory.\cite{readrezayi, mvmr} However, its edge theory\cite{mvmr} is closely related to that of the Abelian $(5,5,3)$ state, which suggests that the HR state may be in the vicinity of the incompressible state and nonetheless affect the physical properties of the system. Recently there have been proposals that compressible states can be molded into incompressible ones.\cite{mjv}

\section{The quantum-well model}\label{quantumwell_section}

So far we have discussed the stability of the one-component Pfaffian state in different finite-width models (Section \ref{pfaffian_section}), and two-component states in a bilayer model where each layer is considered as an infinitely narrow quantum well (Section \ref{twocomponent}). In this Section, we consider an infinite square well of width $w$ in the direction $z \in \left[ 0, w \right]$. The electronic motion in the $z$-direction will then be quantized, yielding an electronic subband structure. 

\subsection{Two-subband approximation}

Instead of a full description with all the electronic subbands, we only consider the two lowest subbands and identify them with the 
two pseudospin states, $\Psi_{\uparrow, \downarrow}=\phi_{\uparrow,\downarrow}(z)Y_{N_\phi/2, N_\phi/2, m}(\theta,\phi)$, where 

\begin{eqnarray}\label{states_qw}
\phi_\uparrow(z) &=& \sqrt{\frac{2}{w}}\sin\left(\frac{\pi z}{w}\right), \label{states_qw_s} \\
 \phi_\downarrow(z) &=& \sqrt{\frac{2}{w}}\sin\left(\frac{2\pi z}{w}\right) \label{states_qw_a},
\end{eqnarray}
and the $Y$'s represent monopole spherical harmonics with $-N_\phi/2 \leq m \leq N_\phi/2$ (we assume that the states are entirely within the lowest LL). We refer to the states (\ref{states_qw_s}) and (\ref{states_qw_a}) as symmetric and antisymmetric, respectively, because of their reflection symmetry with respect to the center of the well. If their energy difference is denoted by $\Delta_{SAS}$, the corresponding second quantized Hamiltonian is given by\cite{albofath}

\begin{eqnarray}\label{qw_hamiltonian}
H &=& - \frac{\Delta_{SAS}}{2} \sum_{m} \left(c_{m\ua}^{ \dagger} c_{m\ua}- c_{m\da}^{\dagger} c_{m\da} \right) \\
\nn
&&+ \frac{1}{2}\sum_{\{ m \}} \sum_{\{ \sigma \}} V_{m_1,m_2,m_3,m_4}^{\sigma_1 \sigma_2 \sigma_3 \sigma_4} c_{m_1\sigma_1}^{\dagger} c_{m_2\sigma_2}^{\dagger} c_{m_4\sigma_4} c_{m_3\sigma_3},
\end{eqnarray}
where $c_{m\sigma}^{(\dagger)}$ annihilates (creates) an electron in the state $m$ with pseudospin $\sigma$.

The matrix elements $V_{m_1,m_2,m_3,m_4}^{\sigma_1,\sigma_2,\sigma_3,\sigma_4}$ can be straightforwardly evaluated from the Haldane pseudopotentials for the resulting in-plane interaction 

\begin{eqnarray}\label{eff_int}
\nonumber V_{\rm 2D}^{\sigma_1,\sigma_2,\sigma_3,\sigma_4}(\vec{r_1}-\vec{r_2})=\\ 
\frac{e^2}{\epsilon l_B} \int dz_1 \int dz_2 \frac{\phi_{\sigma_1}^* (z_1) \phi_{\sigma_2}^* (z_2) \phi_{\sigma_3}(z_1) \phi_{\sigma_4}(z_2)}{\sqrt{(\vec{r_1}-\vec{r_2})^2+(z_1-z_2)^2}},
\end{eqnarray} 
where the position variables are expressed in units of $l_B$ such that the integral is dimensionless.

In this paper we do not make an attempt to quantitatively model the experiment of Ref.~\onlinecite{luhman}, but we are interested in the possible phases that may occur and the transitions between them. Therefore, we expect the model described by the Hamiltonian (\ref{qw_hamiltonian}) to be qualitatively correct and in agreement with other confinement models that assume the lowest subband to be symmetric and the first excited one to have a node in the centre ($z=w/2$).  Any difference of the confining potential away from the infinite square well will modify the energy eigenvalues and the associated wave functions $\phi_\sigma(z)$. However, it is expected that the energies are more strongly affected than the wave functions. In particular, the nodal structure of the wave functions is robust, such that the two lowest eigenstates of the infinite well faithfully represent the underlying features. However, we will allow for the general values of the level splitting $\Delta_{SAS}$ to account for the variations in the eigen-energies.

\subsection{Connection between the quantum-well model and the bilayer Hamiltonian}
\label{sec:connection}

From a more general point of view, the quantum-well model exposed above is a two-component model such as the bilayer model, which
has been used in the discussion of the wide quantum well.\cite{suen} Indeed, the wide quantum well allows the electrons to 
reduce their mutual Coulomb repulsion by exploring more efficiently the $z$-direction, and it has been argued that due to this effect,
a spontaneous bilayer may be formed, under appropriate conditions, in a wide quantum well.\cite{he, suen} Here, a connection is made between both two-component models, on the basis of the Hamiltonian (\ref{qw_hamiltonian}). The intermediate steps in the derivation of the 
effective model may be found in the Appendix A.

The Hamiltonian (\ref{qw_hamiltonian}) may be rewritten in terms of the density and spin-density operators projected to a single
Landau level. The Fourier components of the projected density operator of pseudospin-$\sigma$ electrons reads 
$$
\rhobar_{\sigma}(\bq)=\sum_{m,m'}\left\langle m \left| e^{-i\bq\cdot\bR}\right| m'\right \rangle 
c_{m\sigma}^{\dagger} c_{m'\sigma}\ ,
$$
in terms of the 2D wave vector $\bq$ and the guiding-center operator $\bR$, the latter acting on the states labeled by the quantum numbers $m$. 
It is furthermore useful to define the total (projected) density operator
\beq\label{proj_dens}
\rhobar(\bq)=\rhobar_{\ua}(\bq)+\rhobar_{\da}(\bq)
\eeq
and the projected pseudospin density operators,
\beq\label{proj_spin_dens}
\Sbar^{\mu}(\bq)=\sum_{m,m'}\left\langle m \left| e^{-i\bq\cdot\bR}\right| m'\right \rangle 
c_{m\sigma}^{\dagger} \frac{\tau_{\sigma,\sigma'}^{\mu}}{2} c_{m'\sigma'}\ ,
\eeq
where $\tau_{\sigma,\sigma'}^{\mu}$ are the usual $2\times 2$ Pauli matrices with $\mu=x,y,z$. 

In terms of the projected (pseudospin) density operators, the Hamiltonian (\ref{qw_hamiltonian}) approximately reads
\beqn\label{bilayer_ham}
\nn
H &\simeq& \frac{1}{2}\sum_{\bq} V_{SU(2)}(\bq) \rhobar(-\bq)\rhobar(\bq) \\
  && + 2\sum_{\bq}V_{sb}^x(\bq) \Sbar^x (-\bq)\Sbar^x (\bq)\\
&& - \tilde{\Delta}_{SAS}\Sbar^z(\bq=0)\ ,\nonumber
\eeqn
where the SU(2)-symmetric interaction potential $V_{SU(2)}(\bq)$ and the symmetry-breaking potential $V_{sb}^x(\bq)$ are 
linear combinations of the Fourier-transformed potentials defined in Eq.~(\ref{eff_int}). Their precise form is given in the 
appendix by Eqs. (\ref{sym_int}) and (\ref{sbx_int}), respectively. The Hamiltonian (\ref{bilayer_ham}) neglects a particular
term $\propto \Sbar^z (-\bq)\Sbar^z (\bq)$, which turns out to constitute the lowest energy scale in the interaction Hamiltonian
(\ref{qw_hamiltonian}) [see Eq.~(\ref{en_scales}) in the appendix].

Furthermore,
\beq\label{eff_SAS}
\tilde{\Delta}_{SAS} = \Delta_{SAS} - \gamma \nu \frac{e^2}{\epsilon l_B} \frac{w}{l_B}
\eeq 
is the effective subband gap. The numerical prefactor $\gamma$ depends on the precise nature of the considered confinement potential, and, as shown in the appendix,
the expression (\ref{eff_SAS}) is derived within a mean-field approximation of a particular term in the Hamiltonian 
(\ref{qw_hamiltonian}). The expression (\ref{eff_SAS}) is easy to understand -- whereas the subband gap $\Delta_{SAS}$ 
tends to polarize the system in the $\ua$ state, namely in narrow samples, the second term in Eq.~(\ref{eff_SAS}) 
indicates that the interactions are weaker in the $\da$ subband. From the interaction point of
view, it is therefore energetically favorable to populate the first excited subband. This effect becomes more pronounced in larger
quantum wells. Notice furthermore that this argument also delimits the regime of validity of the two-subband approximation of the
wide quantum well; when the term $\gamma\nu (e^2/\epsilon l_B)\times (w/l_B)$ becomes much larger than the bare
subband gap $\Delta_{SAS}$, the electrons may even populate higher subbands, which are neglected in the present model, and the
system eventually crosses over into a 3D regime.

Notice that the Hamiltonian (\ref{bilayer_ham}) has the same form as the Hamiltonian which describes a 
bilayer quantum Hall system,\cite{moon} up to a rotation from the $z$- to the $x$-axis. 
In this rotated reference frame, one may define the intra- and inter-layer interactions as
\beqn\label{intra}
V_A(\bq) &=& V_{SU(2)}(\bq) + V_{sb}^x(\bq)\\ 
\nn
&=& \frac{1}{4}\left[V_{\rm 2D}^{\ua\ua\ua\ua}(\bq) + V_{\rm 2D}^{\da\da\da\da}(\bq) + 2 V_{\rm 2D}^{\ua\da\ua\da}(\bq)\right] \\ \nn 
 && + V_{\rm 2D}^{\ua\ua\da\da}(\bq)
\eeqn
and 
\beqn\label{inter}
V_E(\bq) &=& V_{SU(2)}(\bq) - V_{sb}^x(\bq)\\ 
\nn
&=& \frac{1}{4}\left[V_{\rm 2D}^{\ua\ua\ua\ua}(\bq) + V_{\rm 2D}^{\da\da\da\da}(\bq) + 2 V_{\rm 2D}^{\ua\da\ua\da}(\bq)\right] \\ \nn
&& -V_{\rm 2D}^{\ua\ua\da\da}(\bq)\ .
\eeqn
As for the case of the true bilayer, the thus defined intra-layer interaction is stronger than the inter-layer interaction, for
all values of $\bq$.

Since our ED calculations employ the Hamiltonian (\ref{qw_hamiltonian}), in order to compare the numerical results with the Halperin states (\ref{halperinwf}) which are the native eigenstates of the true bilayer Hamiltonian (\ref{halperinham}), we can apply the mapping between the two models described above in a reverse fashion. As Halperin wave functions are commonly labeled by the single particle states $|\uparrow \rangle, |\downarrow \rangle$ (which are the eigenstates of $S_z$) and defined by interaction potentials $\{ V_A, V_E \}$, we can imagine a linear transformation (rotation from $z$ to $x$) that transforms them into (unnormalized) symmetric $|+\rangle=|\uparrow \rangle + |\downarrow \rangle $ and antisymmetric $|- \rangle = |\uparrow \rangle - |\downarrow \rangle $ combinations. Then, by inverting the Eqs. (\ref{intra}) and (\ref{inter}), we obtain the set of interaction potentials that generate the Halperin states $(m,m',n)$ in a quantum well description.  In what follows, Halperin states (\ref{halperinwf}) are understood to be indexed by $|+\rangle, |- \rangle$ instead of the usual notation $|\uparrow \rangle, |\downarrow \rangle$, unless explicitly stated otherwise.

\subsection{Energetics of trial wavefunctions}
\label{sec:Monte-Carlo}

To extend the reach of our calculations to system sizes larger than those which can be treated in 
ED, we set up Monte-Carlo simulations of the trial states which have emerged
as good candidates for the ground state. The general strategy of this approach is to obtain an
estimate of the energy in the thermodynamic limit for the different trial states based on a
scaling with system size of their energies. 

As detailed in section \ref{sec:connection} above, we expect formation of two-component wave functions
where $S_x$ is a good quantum number, such that the Halperin wave functions are expressed in terms of the
coordinates of electrons in the $|+\rangle$ and $|-\rangle$ states, 
and lower well, 
indexed below by $\sigma$. We consider cases with equal population of electrons 
in these two bands, or full population of the lowest subband in the ISQW for the single component cases.

In order to calculate efficiently the interaction of electrons in a well of finite width using 
Monte-Carlo simulations, we replace the interaction (\ref{eff_int}) with an effective potential that
reproduces \emph{all} pseudopotential coefficients of the original potential $V_{\rm 2D}$.
Many such potentials can be constructed. Here, we use an interaction of the form proposed 
in Ref.~\onlinecite{JainNoPfaffian}, built from simple polynomials\footnote{In practice, the
use the effective potential (\ref{eq:MC_effective}) is limited to moderately large systems with
$N_\phi\lesssim 60$. For larger systems, it is more suitable to use effective potentials based 
explicitly on the asymptotic behaviour of the in-plane interaction (\ref{eff_int}).}
\begin{equation}
  \label{eq:MC_effective}
  V_{\rm eff}^{\sigma\sigma'}(r) = \sum_{k=-1}^{N_\text{max}^{\sigma\sigma'}} c_k^{\sigma\sigma'} r^k.
\end{equation}
The pseudopotentials of the monomials $r^n$ can be evaluated analytically (generalizing 
Ref.~\onlinecite{FanoOrtolani}). Choosing $c_k$ to match the pseudopotential coefficients of the
interaction (\ref{eff_int}) becomes a simple linear problem. Crucially, we allow for the coefficient of 
the Coulomb term $c_{-1}$ to be varied, also. The number of terms is chosen equal to the minimal 
number required to match the relevant pseudopotentials (odd pseudopotentials $V_{2m+1}$ for intra
(pseudo-)spin interactions and all $N_\phi+1$ terms, otherwise). 

It is habitual in the literature to introduce a neutralizing background, in order to highlight
the correlation energy associated with a wavefunction. We use a background $E_{\rm bg}[\phi]$ 
that matches the distribution $|\phi(z)|^2$ of electrons in their subbands, in order to study 
the \emph{correlation} energy of the different states. However, to establish a final comparison 
between the different wavefunctions, a unique convention for the background is required, and we 
adopt the background of the single layer configuration as a reference point 
$E_{\rm ref} = E_{\rm bg}[\phi_\uparrow]$.

Extrapolation to the thermodynamic limit is undertaken as two separate steps. The correlation energy 
is obtained by linear scaling over the inverse system size $N^{-1}$, using the habitual rescaling of 
the magnetic length $l_B'=[N/(\nu N_\phi)]^{1/2}l_B$.\cite{MorfAmbrumenil} For the two-component
states, the difference in background energy $E_{\rm ref}-\sum_\sigma E
^\sigma_{\rm bg}[\phi_{\sigma'}]$ is
extrapolated separately, and added to the correlation energy.

\section{Competing Phases in the quantum-well model}\label{section_competingphases}

In order to justify the model of the quantum well, in this Section we present the ED study of the Hamiltonian (\ref{qw_hamiltonian}) and analyze the energetics of the relevant trial wavefunctions in Monte-Carlo simulations. We briefly revisit the problem of $\nu=1/2$ extending the results of Refs.~\onlinecite{suen} and \onlinecite{y2} (Section \ref{section_onehalf}) and then present results pertaining to $\nu=1/4$ (Section \ref{section_onefourth}).

\subsection{$\nu=1/2$ in a quantum well}\label{section_onehalf}

At this filling factor the competing phases we consider here are the (3,3,1) Halperin state, the Moore-Read Pfaffian and the HR state. Ref.~\onlinecite{y2} has demonstrated a competition between the multicomponent (3,3,1) state and the fully polarized single-component Moore-Read Pfaffian. In the region of small tunneling, the ground state shows high overlap with the Halperin state; as the tunneling is increased, the Halperin state is destroyed and the Pfaffian takes over. The point of crossover between the two is related to the upward cusp in the activation gap.\cite{y2} 

\begin{figure}[ttt]
\centering
\includegraphics[angle=0,width=\linewidth]{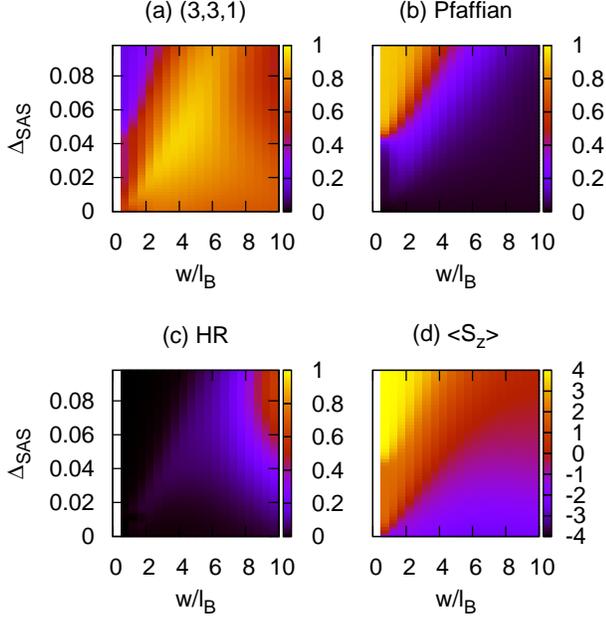}
\caption{(color online) Overlap between the exact Coulomb state of the quantum well for $N=8$ particles at $\nu=1/2$ with the Halperin (3,3,1) state (a),  the Pfaffian (b) and HR (c) states. The expectation value of the $S_z$ component of the pseudospin is plotted in (d).}\label{onehalf_overlap}
\end{figure}

Fig.~\ref{onehalf_overlap} shows our ED results for $8$ particles in the quantum well at the filling factor $\nu=1/2$. Figs. 
\ref{onehalf_overlap}(a), (b), and (c) represent the overlap between the exact ground state and the (3,3,1), the Pfaffian, and the HR
states, respectively, as a function of the well width $w/l_B$ and the bare subband gap $\Delta_{SAS}$. In general, the latter is
a monotonically decreasing function of the well width. Again, we choose $w$ and $\Delta_{SAS}$ as independent parameters of the model. Furthermore, we plot the quantity denoted by $\langle S_z \rangle$, the expectation value of the $S_z= \Sbar^z(\bq=0)$
component of the pseudospin which has the meaning of the ``order parameter" [Fig.~\ref{onehalf_overlap}(d)]. One notices that $\langle S_z\rangle$
continuously crosses over from a full polarization in the $\ua$ subband at low values of $w/l_B$ and a large gap $\Delta_{SAS}$ to 
a polarization in the $\da$ subband for larger quantum wells and small gaps $\Delta_{SAS}$. As it is discussed in the 
previous section, the interactions in a wider quantum well favor a population of the first excited electronic subband $\da$ 
because of the node in the wavefunction in the $z$-direction and, therefore, decrease the effective subband gap.
Indeed, Eq.~(\ref{eff_SAS}) indicates that the crossover line from positive to negative $\tilde{\Delta}_{SAS}$ is characterized
by a border that is linear in $w/l_B$. This behavior is also apparent in Fig.~\ref{onehalf_overlap}(d).  
Notice, however, that for large negative polarizations (large  negative $\tilde{\Delta}_{SAS}$), the two-subband approximation is no  
longer valid and the occupation of even higher electronic subbands  must be taken into account, as already mentioned in Sec. \ref{sec:connection}.

Note, furthermore, that we have defined our (3,3,1) state to be an eigenstate of the $S_x$ operator in the terminology of the true bilayer and not the usual $S_z$ operator (naively defining the Halperin state to be the eigenstate of $S_z$ does not give any appreciable overlap with the exact ground state). There is a simple reason why this needs to be done: because the states of the quantum well possess nodal structure (\ref{states_qw}), the true bilayer states (like the Halperin states) need to be rotated first from $z$ to $x$ direction, in order to match this symmetric/antisymmetric property, before direct comparison can be made.  

With this convention, the (3,3,1) state has its largest overlap ($\lesssim 0.95$) with the exact ground state in the vicinity of the crossover line $\langle S_z\rangle=0$.  However, the overlap remains quite large even in regions beyond this line, where the polarization becomes non-zero [Fig.~\ref{onehalf_overlap}(a)], in agreement with Ref.~\onlinecite{suen}. This behavior may have two different origins. First, one notices that $S_z$ is not a good quantum number if the SU(2) symmetry-breaking terms of the Hamiltonian (\ref{bilayer_ham}) 
in the $x$-direction are taken into  account. Especially in the vicinity of the crossover line $\tilde{\Delta}_{SAS}\simeq 0$, the symmetry breaking is governed by these terms in the $x$-direction, and $\Sbar^x(\bq=0)$, which does not commute with $S_z$, is expected to be a good quantum number. An
alternative origin of the large overlap with the (3,3,1) state even in regions with $\langle S_z\rangle \neq 0$ may be a possible
admixture ($\sim 5\%$) of states to the ground state that are orthogonal the (3,3,1) and possess a finite polarization in the 
$z$-direction.

The largest values of the overlap between the compressible HR state
and the exact ground state are also found in the vicinity of the crossover line from positive to negative $\tilde{\Delta}_{SAS}$,
though at extremely large values of $w/l_B$. Notice that the overlap ($0.64$ for $w/l_B=10.0$) is generally much lower than
for the (3,3,1) state. At large values of the bare subband gap $\Delta_{SAS}$ (and narrow quantum wells), the system becomes polarized
in the $\ua$ subband, and the ground state crosses over smoothly from the (3,3,1) state to the spin-polarized Pfaffian
(overlap of $\lesssim 0.92$). However, the increase in $\Delta_{SAS}$,  somewhat counterintuitively, does not immediately destroy the Halperin state, but at first even increases the overlap.

\begin{figure}[ttt]
\centering
\includegraphics[width=\columnwidth]{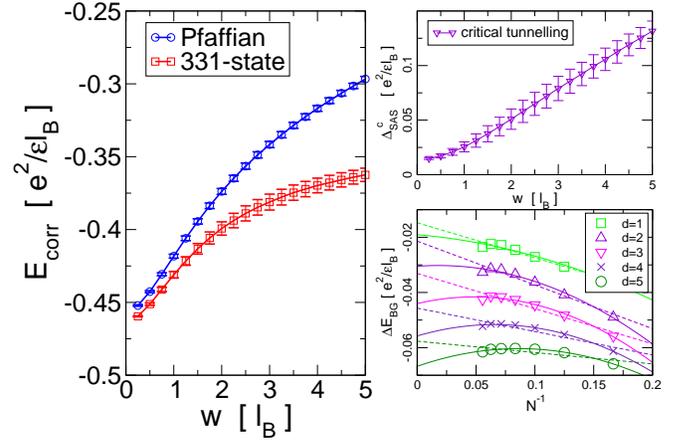}
\caption{(color online) Energies in the thermodynamic limit for the (3,3,1) and Pfaffian states at $\nu=1/2$ (left). Data
  shown is for the infinite square well as a function of the well width $w$. 
  The correlation energies are shown with respect to the single component background. In the absence of tunneling, 
  the (3,3,1)-state has lower energy at all $w$. The critical tunneling strength required to favor the Pfaffian 
  state (top right) and a few typical differences in the extrapolation of the background energies for different values of the well width (bottom right).}\label{onehalf_energetics}
\end{figure}

Finally, Fig.~\ref{onehalf_energetics} shows the results of our Monte-Carlo study of the energies of the (3,3,1) 
and Pfaffian states. The correlation energies of both states were obtained from the finite size scaling of systems
with $N=6$ to $N=18$ electrons as described above in Section \ref{sec:Monte-Carlo}. All data was obtained in 
Monte-Carlo simulations with $10^7$ samples. The uncertainty in the energy of the two-component states was obtained
as the difference between linear and quadratic extrapolation of the background energies (Fig.~\ref{onehalf_energetics}), as this was larger than the
bare numerical errors of the simulation. The energetic competition of
these two phases qualitatively recovers the picture gained from studying the overlaps with the exact ground-state.
Again, some finite amount of tunneling is required for the single component paired state to outcompete the Halperin state. As shown in Fig.~\ref{onehalf_energetics}, the critical tunneling value $\Delta_{SAS}^c$ above which the  Pfaffian state is energetically favored  has a similar upturning shape  
as the boundary of large overlaps for the Pfaffian state in Fig.~\ref{onehalf_overlap}. However, there are some quantitative
differences at small $w$, where the thermodynamic values indicate that polarization occurs at smaller values
of the tunneling.

\subsection{$\nu=1/4$ in a quantum well}\label{section_onefourth}

\begin{figure}[ttt]
\centering
\includegraphics[angle=0,scale=0.4, width=\linewidth]{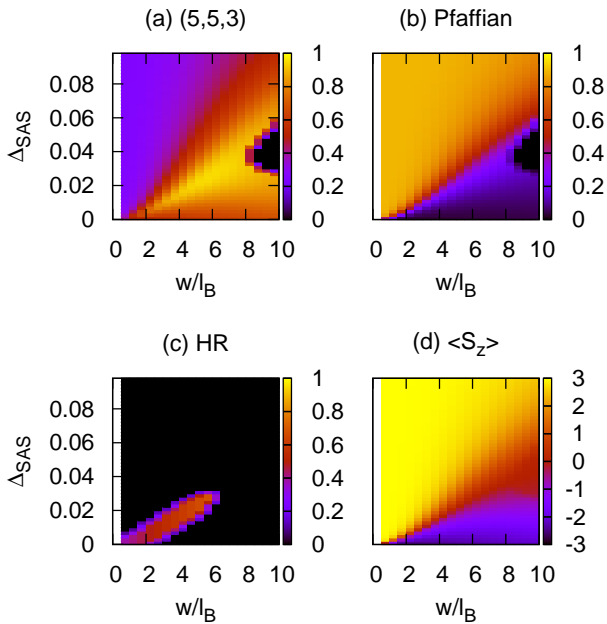}
\caption{(color online) Overlap between the exact Coulomb state of the quantum well for $N=6$ particles at $\nu=1/4$ with the Halperin (5,5,3) state (a),  the Pfaffian (b) and HR (c) states. The expectation value of the $S_z$ component of the pseudospin is plotted in (d).}\label{onefourth_overlap_6}
\end{figure}

\begin{figure}[ttt]
\centering
\includegraphics[angle=0,scale=0.4, width=\linewidth]{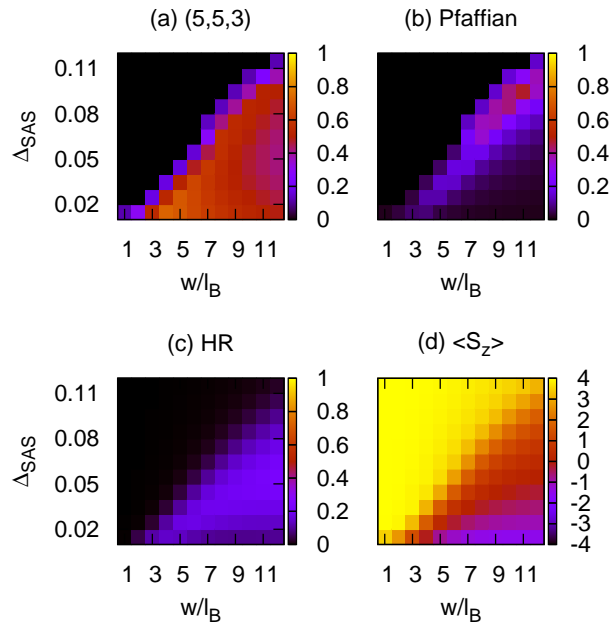}
\caption{(color online) Overlap between the exact Coulomb state of the quantum well for $N=8$ particles at $\nu=1/4$ with the Halperin (5,5,3) state (a),  the Pfaffian (b) and HR (c) states. The expectation value of the $S_z$ component of the pseudospin is plotted in (d).}\label{onefourth_overlap_8}
\end{figure}

\begin{figure}[ttt]
\centering
\includegraphics[angle=270,scale=0.4, width=\linewidth]{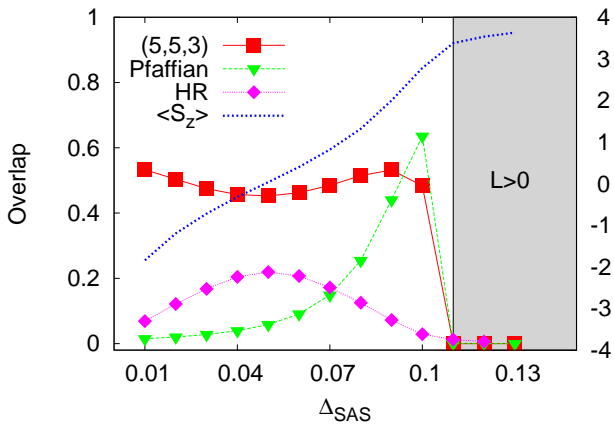}
\caption{(color online) Overlap between the exact Coulomb state of the quantum well for $N=8$ particles at $\nu=1/4$ and $w/l_B=10.5$ with the Halperin (5,5,3) state,  the Pfaffian and HR  states (left axis). The expectation value of the $S_z$ component of the pseudospin is given on the right axis. The shaded region denotes where the ground state is no longer rotationally invariant ($L>0$).}\label{onefourth_crosssection_8}
\end{figure}

\begin{figure}[ttt]
\centering
\includegraphics[width=\columnwidth]{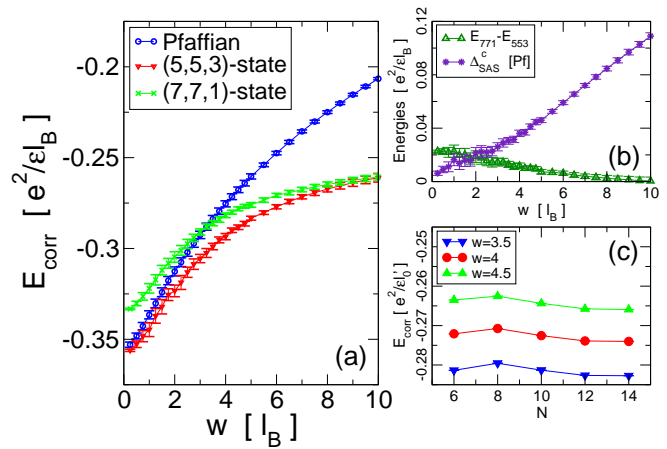}
\caption{(color online) Results from our Monte-Carlo study of states at $\nu=1/4$: (a) correlation energies of the
  Pfaffian-, (5,5,3)- and (7,7,1)-states with respect to the single-component background, as a function of the 
  well-width $w$ in the thermodynamic limit; 
  (b) difference in energy between the different Halperin states and, in particular, the tunneling strength 
  $\Delta_{SAS}^c$ for the Pfaffian state to be favoured over the (5,5,3)-state, and 
  (c) correlation energies for the Pfaffian state in units of rescaled magnetic length\cite{MorfAmbrumenil}
  $l_B'=[N/(\nu N_\phi)]^{1/2}l_B$ for some values of $w$: note the particularly high 
  value at $N=8$.}\label{onefourth_energetics}
\end{figure}

We proceed with analyzing the quantum well at $\nu=1/4$ (Figs. \ref{onefourth_overlap_6} - \ref{onefourth_energetics}). Because of the rapid increase in size of the Hilbert space, there are only two system sizes accessible in ED at this filling factor: $N=6$ and $N=8$. The dimension of the $L_z=0$ sector of the Hilbert space of the latter, taking into account discrete $L_z \rightarrow -L_z$ symmetry, is on the order of $13$ million, thus making $N=6$ the only case amenable to study in great detail. However, for $N=6$  we also must keep in mind the aliasing problem that occurs for (5,5,3) and the Pfaffian (there is no such problem for the HR state). We will present results for both particle numbers because of the important differences between them.  In view of the comments in section \ref{twocomponent}, we note that the overlap with the $(7,7,1)$ state is negligible in the range of widths $w/l_B \lesssim 10.0 $, and therefore we will exclude it from the present discussion of ED results. Note that, similarly to the (3,3,1) state in Sec. \ref{section_onehalf}, the (5,5,3) state hereinafter is defined as an eigenstate of the $S_x$ operator (if defined as an eigenstate of $S_z$, the overlap with the exact ground state is negligible).

In Fig.~\ref{onefourth_overlap_6} we plot the overlap between the ground state of the quantum well for $N=6$ particles at $\nu=1/4$ and the Halperin $(5,5,3)$ state (a), the Pfaffian (b) and the HR state (c), accompanied by the expectation value of the $S_z$ component of the pseudospin (d). These results are reminiscent of  $\nu=1/2$ (Fig.~\ref{onehalf_overlap}); however, due to the smaller energy scale and the gap, it is much easier to polarize the system at $\nu=1/4$. For intermediate values of the width and small tunneling, the maximum overlap with the Halperin $(5,5,3)$ state is high ($0.96$), but the region that would correspond to this phase is quite narrow in comparison to that of $(3,3,1)$. On the other hand, the Pfaffian phase is much more extended. Given the intrinsic tunneling\cite{suen} of the samples, which is of the order of $\Delta_{SAS}/(e^2/\epsilon l_B) \lesssim 0.1$, it seems more likely that the system will be found in this phase than the $(5,5,3)$. 

The small island where the overlap abruptly goes to zero for large $w/l_B$ is due to the ground state belonging to a sector with $L>0$ --- this can be due to the admixture of compressible physics at large widths.  The HR state appears to be present in the transition region between one- and two-component phases, its overlap steadily increasing with $w$ and peaking at $0.7$ for $w/l_B=6.0$.  Because of the fact that the HR state occurs at a different shift on the sphere, we stress that the overlap presented here does not constitute a proof that it is an intermediary phase (moreover, the overlap drops rapidly when larger systems are considered, see Fig.~\ref{onefourth_crosssection_8}).

In Fig.~\ref{onefourth_overlap_8} we plot the same quantities for the system of $N=8$ particles which is expected to display weaker finite-size effects and does not suffer from the aliasing problem. The $(5,5,3)$ state is found in a sizable parameter range, but the maximum overlap is moderate compared to the case previously studied ($0.74$ for $w/l_B=4.5$). While the HR state generally has a small overlap (not exceeding 0.2) and the evolution of  $\langle S_z \rangle$ remains smooth,  the striking difference in comparison with the $N=6$ results (Fig.~\ref{onefourth_overlap_6}) is the Pfaffian phase. Although it similarly develops with the increase in $\Delta_{SAS}$, once the system reaches full polarization, the phase is destroyed.  

To shed more light on how this occurs, it is useful to look at the ``cross section" of Fig.~\ref{onefourth_overlap_8} for a fixed value of the width $w/l_B=10.5$, chosen to represent the region where the Pfaffian phase is most clearly pronounced (Fig.~\ref{onefourth_crosssection_8}). Although the Pfaffian overlap peaks in the region where $(5,5,3)$ starts to drop, very abruptly both overlaps fall to zero, and the ground state is no longer rotationally invariant. The fact that $L>0$ is a hallmark of compressibility. Precisely at the transition point, a small kink is now visible in $\langle S_z \rangle$. The origin of this kink or the reason why the ground state is obtained in $L>0$ sector are not entirely clear at present. However, the zero overlap with the Pfaffian beyond $\Delta_{SAS}/(e^2/\epsilon l_B) = 0.1$ (where the ground state reduces to a spinless case) agrees with our results of the section \ref{pfaffian_section}. Notice that a compressible ground state with $L>0$ may also indicate a phase with modulated charge density,
such as the Wigner crystal. Indeed, an insulating behavior, as one would expect for an electron crystal, has been found at filling
factors slightly above $\nu=1/5$.\cite{WC} Such a state is not captured in the present ED calculations on the sphere, and 
the question whether a Wigner crystal is the true ground state at large values of $\Delta_{SAS}$ in a wide quantum well is beyond the scope of the present paper.

We refer to Monte-Carlo simulations (Fig. \ref{onefourth_energetics}) to obtain additional information about the candidate incompressible states 
from larger model systems. We include systems with $N=6$ to $N=16$ electrons in the finite size scaling for the 
groundstate energies, again using $10^7$ Monte-Carlo samples, and taking errors as the difference between linear 
and quadratic extrapolation of the background energies. The results of this study are summarized in 
Fig.~\ref{onefourth_energetics}, where we compare the Pfaffian to the (5,5,3) and (7,7,1) Halperin wave functions. 
Again, a two-component state is always preferred in the absence of tunneling. At the layer separations shown, 
this is (5,5,3) as shown in Fig.~\ref{onefourth_energetics}(a). These data also confirm that the (7,7,1) state 
becomes relevant only at large well width $w>10l_B$.
In Fig.~\ref{onefourth_energetics}(b), we display the value of tunneling $\Delta_{SAS}^c$ required to polarize
the system into the paired Pfaffian phase. This feature of the energetic competition of (5,5,3) and the Pfaffian
is very close to the results obtained in ED for $N=6$ both qualitatively and quantitatively: the shape of
$\Delta_{SAS}^c(w)$ is nearly linear and reproduces the location where the overlaps with the exact ground-state
cross over between the two trial states, as was shown in Fig.~\ref{onefourth_overlap_6}.
The splitting $\Delta_{SAS}$ required for the Pfaffian to be the ground state is significant, and probably larger
than the splitting in the experiments of Luhman \emph{et al.} \cite{luhman}, which can be estimated to about 
$\Delta_{SAS}\approx 0.069 e^2/\epsilon l_B$ at the sample width $w\approx 10 l_B$ and their baseline electron 
density.

This similarity between the energetics in the thermodynamic limit and the exact spectrum for $N=6$ particles may
be circumstantial. However, there is another indication that the very different behavior at $N=8$ might be
exceptional. In Fig.~\ref{onefourth_energetics}(c), we show the correlation energies of the Pfaffian state 
for different system sizes $N$ and well widths $w$. This representation reveals the case of $N=8$ as having
particularly high energy. This may be a finite-size effect that can be explained in the composite fermion
picture. The Pfaffian wavefunction can be expressed as a paired state of $^4$CF feeling one quantum of
negative effective flux.\cite{ms08,ms05} The shell structure of these composite fermions on the sphere
yields filled shell states for $N=6$ and $N=12$, whereas for $N=8$ two CFs remain in the highest, partially
filled shell. In this configuration, CFs are susceptible to follow Hund's rule by maximising their angular momentum,
breaking rotational invariance. 

For $N=8$ and $N=10$, Hund's rule predicts an angular momentum of $L=4$, which is indeed found
in ED. This gives us confidence that the system is still described by liquid-like composite fermion physics 
at large $\Delta_{SAS}$. We therefore consider the competition between a Hund's rule state and the paired 
Pfaffian state. For a similar situation with weak pairing in a $\nu=1/2+1/2$ bilayer system at large layer
separation, it was argued\cite{ms08} that for larger systems, the shell-filling effects, and Hund's rule, 
should become less important whereas the pairing effects will remain the same strength, as only $\sim \sqrt{N}$ 
benefit from Hund's rule, whereas all ($\sim N$) particles within some gap energy of the Fermi surface 
contribute to pairing. 

Although the above argument speaks in favor of the possibility for a paired Pfaffian state to be realized at $\nu = 1/4$ for large tunneling
gap $\Delta_{SAS}$, we insist on the variational character of the Monte-Carlo calculations. In these calculations, we have indeed
considered several competing candidate wave functions for a liquid ground state at this filling factor. However, this analysis may not
eliminate the possibility that a compressible state, such as that seen in ED, or even other incompressible phases may indeed be singled out as
a true ground state of the system.
 
Finally, we would like to point out that in ED it is possible to calculate the quantity that we refer to as the ``charge gap", 
\begin{equation}\label{chargegap}
\Delta E=E_{N, N_\phi+1}+E_{N, N_\phi-1} - 2E_{N,N_\phi}
\end{equation}
where $E_{N,N_\phi}$ is the ground state energy for a given number of particles $N$ and number of flux quanta $N_\phi$. This quantity probes the response of the system to the introduction of quasiparticles/quasiholes on top of the ground state, and its dependence on $\Delta_{SAS}$ has been used to delineate between the  one-component and two-component phases.\cite{y2} With the appropriate finite-size corrections, Eq.~(\ref{chargegap}) should correspond to the experimentally measurable ``activation" gap\cite{suen} that governs the temperature scaling of longitudinal resistance $R_{xx} \sim \exp{(-\Delta E/2T)}$. For states that undergo a typical one- to two-component transition, such as the one at $\nu=2/3$ (for small tunneling, it is the state of two decoupled Laughlin liquids, $\nu=1/3 + 1/3$, which develops into a single-component $2/3$-state for large tunneling amplitudes\footnote{A single component $\nu=2/3$ state can be regarded either as the particle-hole conjugate of the $\nu=1/3$ Laughlin state, or as composite fermions at negative effective flux filling $p=-2$ CF-LLs.\cite{ms05}}), the charge gap (\ref{chargegap}) displays a minimum as a function of $\Delta_{SAS}$ in the center of the transition region.\cite{suen} On the other hand, for $\nu=1/2$ where the tunneling-driven transition connects the $(3,3,1)$ state and the Pfaffian, the charge gap (\ref{chargegap}) shows an upward cusp. Our calculations of the charge gap (\ref{chargegap}) in the case of $\nu=1/4$ indicate that this quantity is a less robust way to characterize the nature of the ground state than the calculation of the overlaps with trial wave functions. While for $N=6$ particles at $\nu=1/4$ the charge gap displays a minimum as a function of $\Delta_{SAS}$, there is a very weak dependence of $\Delta E$ on $\Delta_{SAS}$ when a larger system of  $N=8$ particles is considered. Thus finite-size effects are too strong in order to extract useful information from Eq.~(\ref{chargegap}) in small systems that can be treated by ED. 

\section{Conclusion}\label{conclusion_section}

In this paper we have presented a systematic study of several candidates for the ground-state wave function at the recently observed\cite{luhman} fraction $\nu=1/4$. Assuming that the (pseudo)spin plays no role, i.e., in a one-component picture, the generalized Moore-Read Pfaffian state (\ref{pfaffian}) shows high overlap for the values of the sample width which are on the order of those in the experiment of Ref.~\onlinecite{luhman}, but only if the confinement in the perpendicular direction is modeled by the ZDS model (\ref{zds_model}). For other confinement models (\ref{isqw}-\ref{fh}) it was not possible to reproduce such high values of the overlap. We believe that this inconsistency means that the high overlap must be due to a special softening of the pseudopotentials that occurs as a pathology of ZDS model but does not appear in other (more realistic) confinement models. 

Therefore, the existence of a fractional quantum Hall state at $\nu=1/4$ is necessarily linked to the specific features of the quantum well used in Ref.~\onlinecite{luhman} that enable the multicomponent physics to manifest itself. Additional degrees of freedom in our theoretical study are conveniently taken into account within the quantum-well model, which is the simplest model that can naturally interpolate between a single layer and bilayer charge distribution as the parameters $w$ and $\Delta_{SAS}$ are varied.  This two-parameter model is related to the previous studies\cite{y2} of the true bilayer with tunneling at $\nu=1/2$ (which had to assume at least three independent parameters) by reproducing the same physical picture of the crossover between the (3,3,1) state and the Pfaffian. 

At the filling factor $\nu=1/4$, we have not been able to produce clear cut evidence for the expected crossover between the (5,5,3) state and the Pfaffian in ED, due to the strong finite-size effects in case of the latter. We have shown that the (5,5,3) state is indeed present for a range of widths and small tunneling gaps $\Delta_{SAS}$, but its maximum overlap is not as high as that of the (3,3,1) state. ED cannot delimit the range of parameters for the Pfaffian phase due to the difference in the results for the two available system sizes, $N=6$ and $N=8$, and the effect of compressible physics which is difficult to treat within the spherical geometry. However, our Monte-Carlo simulations go some way towards clarifying the situation. The correlation energies of the Pfaffian state reveal $N=8$ as a particularly unfavorable system size. We can explain this from the finite-size effect in terms of filling shells of CF orbitals on the sphere. 
The competing $L\neq 0$ states at $N=8$, as well as $N=10$, seem to be related to Hund's rule for CFs. However, the competition between Hund's rule and pairing is likely favorable for the paired state in the thermodynamic limit. In addition, projecting from the two-component model onto the fully polarized (spinless) case, on the other hand, can be seen as analogous to the scenario of LL mixing,\cite{ll_mixing} which may provide another mechanism to stabilize the Pfaffian state via generating three-body terms in the effective interaction. Such effects are beyond the scope of present paper. By analyzing the competition between the paired single component and the Halperin states from their variational wave functions, we find, in the Monte-Carlo simulations, that the tunneling gap $\Delta_{SAS}^c$ required to form a single-component state roughly behaves linearly as $1.0\times (w/l_B) \times 10^{-2} e^2/\epsilon l_B$.
Although the tunneling splitting indicated for the experiment described in Ref.~\onlinecite{luhman} is not far below 
the transition between the Pfaffian and (5,5,3), our numerics still show it safely in the two-component regime 
of the (5,5,3)-wavefunction. 

Although we believe that our quantum well model takes properly into account the effects of finite thickness, we have entirely neglected the effect of the in-plane magnetic field which may nevertheless prove essential in order to stabilize the incompressible state at $\nu=1/4$. The existing experimental work\cite{lay} on the $\nu=2/3$ state witnessed that the introduction of an in-plane magnetic field may lead to a strengthening of the minimum in $R_{xx}$, thus inducing the same one-component to two-component transition as by varying $\Delta_{SAS}$. Similar strengthening occurs for $\nu=1/2$ if the tilt is not too large.\cite{lay} Therefore, the application of the in-plane field may be a likely reason to further stabilize the (5,5,3) state at $\nu=1/4$ if the symmetric-antisymmetric gap $\Delta_{SAS}$ is sufficiently small. However, Ref.~\onlinecite{luhman} also pointed out the difference between $\nu=1/2$ state and $\nu=1/4$ state: when the electron density is increased, the former displays a deeper minimum in $R_{xx}$ while the latter remains largely unaffected. This difference suggests that in the case of $\nu=1/4$ the quantum-well ground state may be effectively fully polarized and in the class of the Pfaffian rather than the two-component, (5,5,3) state.  

In order to answer without ambiguity which of the two possibilities is actually realized in the quantum well under the experimental conditions of Ref.~\onlinecite{luhman}, it would be useful to know the dependence of the activation gap as a function of $\Delta_{SAS}$ and also as a function of transferred charge from the front to the back of the quantum well using a gate biasing. These results would help to discriminate between the one-component and two-component nature of the ground state.

\section*{Acknowledgments}

This work was funded by the Agence Nationale de la Recherche under Grant No. ANR-JCJC-0003-01. 
ZP was supported by the European Commission, through a Marie Curie Foundation
contract MEST CT 2004-51-4307 and Center of Excellence Grant CX-CMCS. MVM  was supported by the Serbian 
Ministry of Science under Grant No. 141035. GM would like to thank Steven Simon for stimulating discussions.

\appendix\label{app:A}
\section{Effective bilayer description of the wide quantum well}

As in Sec. \ref{quantumwell_section}, we consider
the quantum well to be symmetric around $w/2$, i.e. the lowest subband ($\ua$) state is symmetric, and the first excited one
($\da$) is antisymmetric. Furthermore, we consider, in this section, the electrons to be in the 2D plane, for illustration reasons, 
although the conclusions remain valid also in the spherical geometry. In this appendix, we yield the derivation of the effective
bilayer description of the wide quantum well. 

The interaction part of the Hamiltonian (\ref{qw_hamiltonian}) consists of a density-density interaction and terms beyond, which may
be described as a spin-spin interaction. Indeed, the density-density part consists of the effective interactions (\ref{eff_int})
$V_{\rm 2D}^{\ua\ua\ua\ua}$, $V_{\rm 2D}^{\da\da\da\da}$, and $V_{\rm 2D}^{\ua\da\ua\da}=V_{\rm 2D}^{\da\ua\da\ua}$. Notice that the interactions
in the first excited subband ($\da$) are generally weaker than in the lowest one ($\da$) because the wave function (\ref{states_qw_a})
$\phi_{\da}(z)$ possesses a node at $w/2$, in the center of the well, i.e. $V_{\rm 2D}^{\ua\ua\ua\ua} > V_{\rm 2D}^{\da\da\da\da}$. With the
help of the (spin) density operators (\ref{proj_dens}) and (\ref{proj_spin_dens}), 
the density-density part of the interaction Hamiltonian reads
\beqn\label{H_dens_dens}
\nn
H &=& \frac{1}{2}\sum_{\bq} V_{SU(2)}(\bq) \rhobar(-\bq)\rhobar(\bq) \nn \\
 && + 2\sum_{\bq}V_{sb}^z(\bq) \Sbar^z (-\bq)\Sbar^z(\bq)\\
&& + \sum_{\bq}V_{B}^z(\bq) \rhobar (-\bq)\Sbar^z(\bq)\ , \nn
\eeqn
in terms of the SU(2)-symmetric interaction
\beq\label{sym_int}
V_{SU(2)} (\bq)= \frac{1}{4}\left[V_{\rm 2D}^{\ua\ua\ua\ua}(\bq) + V_{\rm 2D}^{\da\da\da\da}(\bq) + 2 V_{\rm 2D}^{\ua\da\ua\da}(\bq)\right],
\eeq
and the SU(2)-symmetry breaking interaction terms
\beq\label{sbz_int}
V_{sb}^z (\bq)= \frac{1}{4}\left[V_{\rm 2D}^{\ua\ua\ua\ua}(\bq) + V_{\rm 2D}^{\da\da\da\da}(\bq) - 2 V_{\rm 2D}^{\ua\da\ua\da}(\bq)\right]
\eeq
and 
\beq\label{Bz_int}
V_{B}^z (\bq)= \frac{1}{2}\left[V_{\rm 2D}^{\ua\ua\ua\ua}(\bq) - V_{\rm 2D}^{\da\da\da\da}(\bq) \right]\, .
\eeq

The remaining 12 interaction terms, which may not be treated as density-density interactions, fall into two different classes; 
the 8 terms with three equal spin orientations $\sigma$ and one opposite $-\sigma$ are zero due to the antisymmetry of the integrand
in Eq.~(\ref{eff_int}). The remaining 4 interaction terms with two $\ua$-spins and two $\da$-spins are all equal due to the symmetry 
of the quantum well around $w/2$, 
\beq\label{sbx_int}
V_{sb}^x\equiv V_{\rm 2D}^{\ua\ua\da\da}=V_{\rm 2D}^{\da\da\ua\ua}=V_{\rm 2D}^{\ua\da\da\ua}=V_{\rm 2D}^{\da\ua\ua\da} .
\eeq 
They yield
the term
\beq\label{ham_sbx}
H_{sb}^z = 2\sum_{\bq}V_{sb}^x(\bq) \Sbar^x (-\bq)\Sbar^x (\bq)\, ,
\eeq
which needs to be added to the interaction Hamiltonian (\ref{H_dens_dens}), as well as the term
\beq\label{SAS}
H_{SAS}=-\Delta_{SAS}\, \Sbar^z(\bq=0)\, ,
\eeq
which accounts for the electronic subband gap between the $\ua$ and the $\da$ levels.

Collecting all terms, the Hamiltonian (\ref{qw_hamiltonian}) thus becomes
\begin{widetext}
\beqn\label{ham_final}
\nn
H &=&  \frac{1}{2}\sum_{\bq} V_{SU(2)}(\bq) \rhobar(-\bq)\rhobar(\bq) + 2\sum_{\bq}V_{sb}^x(\bq) \Sbar^x (-\bq)\Sbar^x (\bq)
+ 2\sum_{\bq}V_{sb}^z(\bq) \Sbar^z (-\bq)\Sbar^z(\bq) \\
&&+ \sum_{\bq}V_{B}^z(\bq) \rhobar (-\bq)\Sbar^z(\bq)  -\Delta_{SAS}\, \Sbar^z(\bq=0)\ .
\eeqn
\end{widetext}
Several comments are to be made with respect to this result. First, we have checked that for the infinite-square-well model as
well as for a model with a parabolic confinement potential there is a natural hierarchy of the energy scales in the Hamiltonian 
(\ref{ham_final}),  
\beq\label{en_scales}
V_{SU(2)}\ >\ V_{sb}^x\ \gtrsim\ V_{B}^z\ \gtrsim\ V_{sb}^z\ . 
\eeq
This hierarchy is valid both for the interaction potentials in Fourier space as for the pseudopotentials. 

Whereas the first term of the Hamiltonian describes the SU(2)-symmetric interaction, the second and the third one break this SU(2) symmetry. Because $V_{sb}^x(\bq) > V_{sb}^z(\bq) > 0$ for all values of $\bq$, states
with no polarization in the $x$- and the $z$-direction are favored, with $\langle S^x\rangle = 0$ and $\langle S^z\rangle = 0$,
respectively. Due to the hierarchy (\ref{en_scales}) of energy scales, a depolarization in the $x$-direction is more relevant than
that in the $z$-direction. These terms are similar to those one encounters in the case of a bilayer quantum Hall system, where due
to the finite layer separation a polarization of the layer isospin in the $z$-direction costs capacitive energy.\cite{moon}

The fourth term of the Hamiltonian (\ref{ham_final}) is due to the stronger electron-electron repulsion in the lowest electronic
subband as compared to the first excited one, where the wave function possesses a node at $z=w/2$. In order to visualize its effect,
one may treat the density, which we consider to be homogeneous in an incompressible state, on the mean-field level, 
$\langle \rhobar(\bq)\rangle = \nu \delta_{\bq,0}$, in which case the fourth term of Eq.~(\ref{ham_final}) becomes
$\nu V_B^z(\bq=0)\Sbar^z(\bq=0)$ and, thus, has the same form as the subband-gap term (\ref{SAS}). It therefore renormalizes
the energy gap between the lowest and the first excited electronic subbands, and it is natural to define the effective 
subband gap as 
\beqn\label{eff_SAS_A}
\nn
\Delta_{SAS}\rightarrow \tilde{\Delta}_{SAS} &=&\Delta_{SAS} - \nu V_B^z(\bq=0)\\
&=& \Delta_{SAS} - \gamma \nu \frac{e^2}{\epsilon l_B} \frac{w}{l_B}\ ,
\eeqn
where $\gamma$ is a numerical prefactor that depends on the precise nature of the well model.

\end{document}